\documentclass[conference]{IEEEtran}
\IEEEoverridecommandlockouts
\usepackage{cite}
\usepackage{amsmath,amssymb,amsfonts}
\usepackage{algorithmic}
\usepackage{graphicx}
\usepackage{siunitx}
\usepackage{textcomp}
\usepackage[nolist]{acronym}
\def\BibTeX{{\rm B\kern-.05em{\sc i\kern-.025em b}\kern-.08em
    T\kern-.1667em\lower.7ex\hbox{E}\kern-.125emX}}

\begin{document}

\begin{acronym}
\acro{ids}[IDS]{Intrusion Detection System}
\acro{ips}[IPS]{Intrusion Prevention System}
\acro{it}[IT]{Information Technology}
\acro{ipc}[IPC]{Industrial Personal Computer}
\acro{plc}[PLC]{Programmable Logic Controller}
\acro{hmm}[HMM]{Hidden Markov-Model}
\acro{cps}[CPS]{Cyber Physical System}
\acro{siem}[SIEM]{Security Information and Event Management}
\acro{scada}[SCADA]{Supervisory Control And Data Acquisition}
\acro{mes}[MES]{Manufacturing Execution System}
\acro{erp}[ERP]{Enterprise Resource Planning}
\acro{roc}[ROC]{Receiver Operating Characteristic}
\acro{oisf}[OISF]{Open Information Security Foundation}
\end{acronym}

\title{A Question of Context:\\Enhancing Intrusion Detection by Providing Context Information
\thanks{This is a preprint of a work published at the 2017 Internet of Things Business Models, Users, and Networks.
Please cite as follows:
S. D. Duque Anton, D. Fraunholz, S. Teuber, and H. D. Schotten, : ``A question of context: Enhancing intrusion detection by providing context information.'' In: 2017 Internet of Things Business Models, Users, and Networks. IEEE, 2017. pp. 1-8
}
}

\author{\IEEEauthorblockN{1\textsuperscript{st} Simon Duque Anton, 2\textsuperscript{nd} Daniel Fraunholz, \\ 4\textsuperscript{th} Hans D. Schotten}
\IEEEauthorblockA{\textit{Intelligent Networks Research Group} \\
\textit{German Research Center for Artificial Intelligence}\\
Kaiserslautern, Germany\\
\{simon.duque\_anton, daniel.fraunholz,\\hans\_dieter.schotten\}@dfki.de}
\and
\IEEEauthorblockN{3\textsuperscript{rd} Stephan Teuber}
% SDA: IMHO, ``Information Security Group'' sounds better, but it is probably a fixed term
\IEEEauthorblockA{\textit{Group Information Security} \\
\textit{Volkswagen AG}\\
Wolfsburg, Germany\\
stephan.teuber@volkswagen.de}
}

\maketitle

\begin{abstract}
Due to the fourth industrial revolution,
and the resulting increase in interconnectivity,
industrial networks are more and more opened to publicly available networks.
Apart from the huge benefit in manageability and flexibility,
the openness also results in a larger attack surface for malicious adversaries.
In comparison to office environments,
industrial networks have very high volumes of data.
In addition to that,
every delay will most likely lead to loss of revenue.
Hence,
intrusion detection systems for industrial applications have different requirements than office-based intrusion detection systems.
On the other hand,
industrial networks are able to provide a lot of contextual information due to manufacturing execution systems and enterprise resource planning.
Additionally,
industrial networks tend to be more uniform,
making it easier to determine outliers.
In this work,
an abstract simulation of industrial network behaviour is created.
Malicious actions are introduced into a set of sequences of valid behaviour.
Finally,
a context-based and context-less intrusion detection system is used to find the attacks.
The results are compared and commented.
It can be seen that context information can help in identifying malicious actions more reliable than intrusion detection with only one source of information,
e.g. the network.
\end{abstract}

\begin{IEEEkeywords}
Context-aware anomaly detection, Intrusion Detection System, Simulation, IT-Security, Industrial Application
\end{IEEEkeywords}

\section{Introduction}
\label{sec:intro}
 The increasing interconnectivity and the opening of industrial networks to the outside creates the need for security measures.
Such measures,
e.g. firewall, antivirus software or \ac{ids}, 
have been used in office \ac{it} for a long time.
They are,
however,
new to industrial applications.
Security was not an issue in industry for two reasons~\cite{Igure.2006}:
First, 
\ac{scada} systems were physically separated from the internet.
And second,
each system is unique due to its singular purpose,
making it hard for an attacker to exploit it.
Both assumptions hold no longer true as the recent and not-so-recent spectacularly show.
% TODO: Cite our great SCADA-paper? If it is accepted...
This lead to the drastic increase in software and network security for industrial applications.
Since industrial networks are different in structure and purpose than office networks,
the same solutions cannot be transferred readily.
Instead,
new security solutions have to be developed.
Industrial networks are highly specialised,
% Kernanahme unseres Modells -> später bezug drauf nehmen
creating unique but very repetitive traffic~\cite{Zhou.2011}.
Furthermore,
industrial networks are connected to \acp{cps},
meaning they are intended to interact with the real world.
And lastly,
most protocols in the industrial environment do not employ security mechanisms such as authentication and encryption\cite{Zhu.2011, Porros.2010},
as is standard in the office world.
\\ \par
This work is structured as follows.
In section~\ref{sec:sota}, the state of the art is described.
After that, the use case that is considered in this work is introduced in section~\ref{sec:uc_descr}.
The corresponding formal model is derived in section~\ref{sec:formal_model}.
In section~\ref{sec:analysis}, the analysis is conducted and the results are presented.
A conclusion is drawn in section~\ref{sec:conclusion}.

\section{State of the Art}
\label{sec:sota}

There are many different approaches for \acp{ids};
Luh et al. categorized them in their survey \textit{Semantics-aware detection of targeted attacks} into different groups \cite{Luh.2017}.
They identified the main categories ``Host-based'',
``Network-based'' and ``Multi-source'' with several sub-categories.
As we focus our work on industry networks and their components,
we will not focus on individual \acp{ids} for specific hosts,
but on Multi-source and Network-based approaches.
One popular implementation of a Network-based \ac{ids} is \textit{Snort} by Cisco \cite{Roesch.1999},
which has grown from a lightweight \ac{ids} to a tool that can be used as a full-grown network-based \ac{ips}.
The signature-based approach utilizes user contributed rules to catch different specific instances of network attacks.
Another well known network monitoring framework,
which can be used as a network \ac{ids},
is \textit{Bro} \cite{Paxson.1999}.
Bro is split into layers,
where the ``Event Engine'' performs several integrity checks before it handles the packets sent between senders an recievers that are organised as touples -- Bro decides if it should return the whole packet,
just the header or nothing at all,
depending on the used protocol.
The ``Policy Script Interpreter''-layer is used to check if the handled packets generated any events that were specified before and then decides if a Bro script should be executed.
These scripts can include several tasks like generating new events,
logging functions and modifying internal states.
There are more popular open source network \acp{ids} like \textit{Suricata},
which was first released in 2010 by the \ac{oisf} \cite{JoshuaS.White.2013},
and \textit{Kismet},
which aims at wireless networks \cite{Hindarto.2010}.
% SDA: IM not so HO, a new paragraph looks nicer
\\ \par
According to Jyothsna et al.,
another method to categorize \acp{ids} is by classifying them into ``signature based detection'' and ``anomaly based detection'' \cite{Jyothsna.2011}.
While the first group looks for patterns of known attacks and compares them with the active system,
the latter category -- the anomaly based approach -- builds a model of the normal behaviour and matches this learned behaviour with the running system.
While the first approach can't find novel attacks,
the second one needs training on a normal behaving system.
According to Thames and Schaefer, 
the requirements for security aspects differ in many points for \ac{cps} and regular IT systems \cite{Thames.2017}.
As we focus primarily on industry networks with heterogeneous participants like \acp{plc},
\acp{ipc} and \acp{cps} with different operating systems,
used protocols and various available resources,
we have to use a more adjustable approach.
\\ \par
To address these problems we used an approach with \acp{hmm} that uses context information.
An approach for the usage of Markov chains was published by Ye \cite{Ye.2000}.
In her paper \textit{A Markov Chain Model of Temporal Behavior for Anomaly Detection} she introduced her technique where the system learns from historic data to distinguish between normal behavior and intrusive activities.
% TODO: check
Due further developed that approach in \cite{Du.2004} to a \ac{hmm} \ac{ids} to learn patterns of Unix processes.
Hu et al. introduced an \ac{ids} that uses system call based \acp{hmm} \cite{Hu.2009}.
They focussed on reducing the amount of submodels required to model the scenario and the impact of pre-processing data on training time.
Yola{\c{c}}an suggested in his approach to enrich the data with context information~\cite{Yolacan.2016}.
\\ \par
The above mentioned \acp{ids} are host-based and do not focus on other available data sources like network traffic.
More recent research also show first advances in the field of \acp{hmm} in network-based intrusion detection.
Chen proposed a classification method for detecting attacks in different attack stages within regular IT\ networks~\cite{Chen.2016}.
Zohrevand used \acp{hmm} to develop a system for anomaly detection in water supply \ac{scada} systems showing that their approach outperforms other contestants making \acp{hmm} a feasible approach also for industrial networks\cite{Zohrevand.2016}.

\section{Use Case Description}
\label{sec:uc_descr}
There are numerous possible attack vectors on industrial applications.
As described in section~\ref{sec:intro},
each company's network is different and therefore a wholesome security solution is infeasible.
As we described in previous works,
there are, 
however,
common attack motivations and attacker types~\cite{Fraunholz.2017}.
Some attackers aim at extracting information,
so-called ``espionage''.
Others try to cause harm,
in covert operations or in the open,
so-called ``sabotage''.
Not only the goals,
but also the means of achieving them are numerous and strongly depend on the use case.
Therefore,
creating a model that covers all possible attacker scenarios is impracticable.
Instead,
we focus on a subset of all attacks and describe a special manifestation of sabotage.
We consider an attack scenario where an attacker reprograms a \ac{plc}.
This reprogramming leads to unexpected behaviour of the \ac{cps} with the potential to cause harm and damage.
In our opinion,
there are two scenarios how the attacker is able to reprogram the \ac{plc},
and four possible ways to detect this attack.
An attacker can either
\begin{itemize}
\item take over a valid programming device and use it maliciously or
\item introduce a malicious device to the network.
\end{itemize}
How she circumvents the perimeter is beyond the scope of this work,
there are several related works discussing physical and network security~\cite{Northcutt.2005}.
Furthermore,
the given attack can be detected in one of four ways.
\paragraph{Identity}
If each programming device and \ac{plc} is provided with a cryptographically secure identity,
the adversary has to break it.
With state of the art authentication, 
this can be considered impossible.
However,
most industrial applications rarely use authentication on devices~\cite{Zhu.2011}.
This detection mechanism only discovers attacks by an attacker that introduced a malicious device.
If malicious programming is executed with a valid device, 
a valid identity is provided.

\paragraph{Statistically}
Depending on the way an attacker reprograms devices,
statistical deviations from normal behaviour could be observed.
This is possible for both ways an attacker can modify the system as described above,
but only if the attacker significantly differs from the standard system behaviour.
A Dolev-Yao intruder model~\cite{Dolev.1983} assumes the attacker possesses knowledge of the system,
enabling her to perfectly mimic the standard behaviour.

\paragraph{Payload-based}
%TODO: VERSTEHT MAN DEN ABSATZ?
Deep packet inspection is employed by several \ac{ids} to discover content that is capable of causing harm to a system.
In industrial applications,
a complex model of the production facility is necessary to determine which parameters can cause damage.
Even more,
a value that is well within safety boundaries can still lead to the production of goods that are low quality.
Therefore,
it is difficult to determine malicious traffic by payload in industrial applications.

\paragraph{Context-based}
Sometimes,
information about network traffic alone is not enough to discover attacks.
In this case,
adding context information can enrich the data source and help to provide a broader view on a system.
This principle is employed by \ac{siem} systems.
% TODO: reference here our context-paper?
\\ \par
In this work,
we assume that no identity information is provided,
allowing an attacker to spoof entities in the network.
Furthermore,
we assume she does not strictly stick to the standard system behaviour,
creating statistical deviations that could be detectable.
We then compare the statistical analysis with context-based analysis.

\section{Formal Model}
\label{sec:formal_model}
For analysis,
a formal model was created.
The formal model was used to generate data,
as well as to analyse the data with respect to anomalies induced by attacks.
The model consists of three parts:
\begin{itemize}
\item Products
\item \acp{ipc}
\item \acp{plc}
\end{itemize}
The products symbolise the intended output of a fictional production process.
Depending on the change of product, 
several \acp{ipc} have to be reconfigured.
These \acp{ipc} that are reconfigured are a subset of all available \acp{ipc}.
Each \ac{ipc} controls four \acp{plc} that are reprogrammed in case of reconfiguration.
The order of reprogramming the \acp{plc} follows a probability distribution defined by a \ac{hmm}~\cite{Baum.1966}.
A \ac{hmm} $\lambda$ is defined in equation~\ref{eq:hmm}.

\begin{equation}
\label{eq:hmm}
\lambda = (S, V, A, B, \pi)
\end{equation}

\begin{itemize}
\item $S = \{PLC_{1, 1}, ..., PLC_{i, j}\}$, 
\item $V = id(S)$,
\item $A \in \mathbb{R}^{(i \cdot j) \times (i \cdot j)}$, the state transition matrix,
\item $B \in \mathbb{R}^{(i \cdot j) \times (i \cdot j)}$, the observation matrix,
\item $\pi \in \mathbb{R}^{(i \cdot j)}$, the starting distribution describes the probability $\pi_{n} = P(X_{1} = s_{n})$ of the first state to be $s_{n}$
\end{itemize}

$i$ is the Number of \acp{ipc} and $j$ the Number of \acp{plc} per \ac{ipc}.
 $id()$ is a function that maps a \ac{plc} to a unique identifier as each \ac{plc} only outputs its identifier.

%TODO: insert: limitations and explain (with formula) that the Markov-model only considers the last state
% ST: i think that this todo is already taken care of in sec:analysis
% SDA: I think you're right :-) I just forgot to remove this

\section{Analysis and Results}
\label{sec:analysis}

\begin{figure}[htbp]
\centerline{\includegraphics[width=0.45\textwidth]{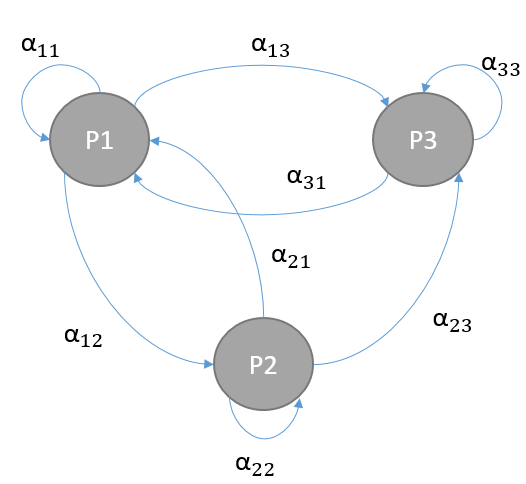}}
\caption{Product Transitions}
\label{fig:prod_trans}
\end{figure}

The model introduced in section~\ref{sec:formal_model} was implemented with specific parameters:
\begin{itemize}
\item $3$ products being produced by
\item $5$ \acp{ipc} controlling
\item $4$ \acp{plc} each
\end{itemize}
The products are chosen randomly,
following a Markov process~\cite{Eberle.2015}.
The possible transitions are shown in figure~\ref{fig:prod_trans},
the corresponding transition probabilities are shown in table~\ref{tab:trans_probs}.
In total, 
24 different setups were simulated and analysed.
These setups consist of two ways the attacker reprograms the \acp{plc},
three different numbers of malicious actions introduced,
and four ways the \acp{ipc} are reconfigured and reprogram the \acp{plc} due to the change of product.
The simulation setups are an abstraction of real-world behaviour.
As explained in section~\ref{sec:uc_descr}, 
an operator knowing every transition of her system would be able to detect any attack,
unless the attacker also knew the transitions and was able to mimic this behaviour.
In this case, 
the operator could only tell valid from malicious attacks by consulting the necessity of changing states due to changes in production.
\\ \par
The goal of this simulation is to derive the importance of context information for detection of various malicious traffic in different kinds of valid traffic.
We first created a set of sequences of valid actions and transitions.
The transitions were saved in one,
respectively three matrices.
The first matrix contained all transitions, 
regardless of the product to be produced.
The three matrices contained the transitions for the correlating product,
introducing the notion of context.
In the real world,
this kind of information is easily accessible via \ac{mes} and \ac{erp},
as,
for example,
proposed in~\cite{Duqueanton.2017}.
First,
the matrices were initialised by \num{10000} cycles of product change.
Normalisation was applied to the matrices.
Then we inserted first one,
then two and finally four malicious packets into each sequence of the set.
After that,
the transition probabilities of the malicious actions were calculated.
% ST: extracted?
% SDA: Jap, there are now malicious parts in the valid actions.
% SDA: Therefore, there are new transitions, from malicious to valid and vice versa.
% SDA: Those transition probabilities are calculated as they will be used to determine the maliciousness.
The probabilities of the attacks,
as well as the valid actions,
were compared to a threshold varying from $0$ to $1$ in steps of $0.05$.
% ST: looks more like 0.05 to me?
% SDA: Jap, of course
This allowed for the calculation of \acp{roc},
as well as the f-measure.
The f-measure,
or F1-score,
 is a metric to determine the quality of a classifier~\cite{Rijsbergen.1979},
% ST: classifier?
% SDA: Sounds better
as written in equation~\ref{eq:f-measure}.

\begin{equation}
\begin{split}
\label{eq:f-measure}
F_{1} = 2 \cdot \dfrac{precision \cdot recall}{precision + recall} \\
precision = \dfrac{t_{p}}{t_{p}+f_{p}} \\
recall = \dfrac{t_{p}}{t_{p}+f_{n}}
\end{split}
\end{equation}

One of the most noteworthy properties of a Markov model is its inability to contain memory of states before the previous one,
the so-called \textit{Markov property}.
This means that each transition probability only depends on the previous state,
completely independent of its predecessors before the current state.
This behaviour is noted mathematically in equation~\ref{eq:markov}.
The transition probabilities of each state transition only consider the previous state.
where $p$ is a matrix of transition probabilities,
$P$ the probability of an event,
$X$ an event,
$s$ a state and $t$ a time step.

\begin{equation}
\begin{split}
\label{eq:markov}
p_{i, j} = P(X_{t+1} = s_{j} | X_{t} = s_{i}) \\
i, j = 1, ..., m
\end{split}
\end{equation}

Each sequence of malicious programming actions was considered an attack and should have been found.
It was,
however,
sufficient to detect one malicious segment within a sequence.
In reality,
finding only one of a series of malicious programming actions would be sufficient to raise awareness for an attack.
First, 
we analyse attacks that are introduced to several scenarios in static order in subsection~\ref{ssec:aiso}.
Then,
attacks that are introduced into the same scenarios in different order are examined in subsection~\ref{ssec:airo}.
The results are presented in subsection~\ref{ssec:results}.

\subsection{Attacks in Static Order}
\label{ssec:aiso}

In these setups,
the attacks always were in a static order.
In turn,
one,
two and four attacks were introduced into four different kinds of valid actions.

\begin{figure*}[htbp]
\centerline{\includegraphics[width=0.9\textwidth]{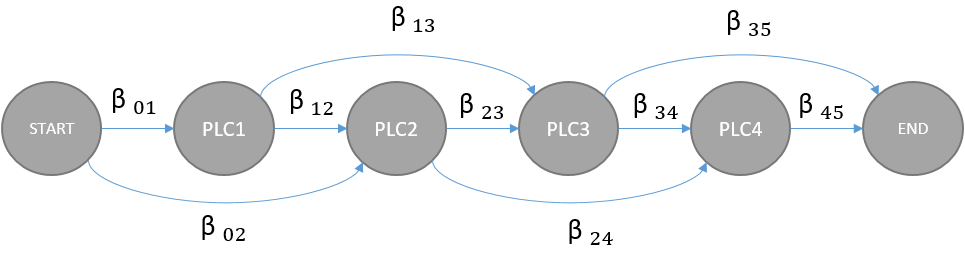}}
\caption{PLC Transitions}
\label{fig:plc_trans}
\end{figure*}

\paragraph{Setup 1}
The \acp{ipc} that are reconfigured for each product are shown in table~\ref{tab:ipc_prods}.
If a product is selected,
the \acp{ipc} are reconfigured in their numerical order.
The products are randomly chosen with probabilities as listed in table~\ref{tab:trans_probs},
Upon reconfiguration,
each \ac{ipc} reprograms its corresponding \acp{plc}.
This reprogramming process is depicted in figure~\ref{fig:plc_trans}, 
the according transition probabilities are listed in table~\ref{tab:plc_probs}.
The given probability distribution allows for some \acp{plc} to not be reprogrammed.
This behaviour is supposed to create a small uncertainty that,
in reality,
could arise from maintenance or reprogramming steps that were undergone.
Still,
one major difference between office- and industrial-\ac{it} lies in the repetitive traffic as described in section~\ref{sec:intro}.
This repetitive behaviour is modelled by reconfiguring the same \acp{ipc} in the same order for each product and the order of reprograming of the \acp{plc}.

\begin{table}
\centering
\caption{\acp{ipc} that are programmed for each individual product}
\label{tab:ipc_prods}
\begin{tabular}{| l | c | c | c |}
\hline
 & Product 1 & Product 2 & Product 3 \\
\hline
IPC 1 & x & & x \\
\hline
IPC 2 & & x & x \\
\hline
IPC 3 & x & x & \\
\hline
IPC 4 & & & x \\
\hline
IPC 5 & x & & x \\
\hline
\end{tabular}
\end{table}

\begin{table}
\centering
\caption{Transition Probablities for the Products}
\label{tab:trans_probs}
\begin{tabular}{| l | c | c | c |}
\hline
 & Product 1 & Product 2 & Product 3 \\
\hline
 Product 1 & 0.6 & 0.2 & 0.2 \\
\hline
Product 2 & 0.4 & 0.6 & 0 \\
\hline
Product 3 & 0.6 & 0.2 & 0.2 \\
\hline
\end{tabular}
\end{table}

\begin{table}
\centering
\caption{Starting Probabilities for the Products}
\label{tab:starting_probs}
\begin{tabular}{| c | c | c |}
\hline
Product 1 & Product 2 & Product 3 \\
\hline
0.8 & 0.1 & 0.1 \\
\hline
\end{tabular}
\end{table}

\begin{table}
\centering
\caption{Transition Probabilities for the \acp{plc}}

\label{tab:plc_probs}
\begin{tabular}{| l | c | c | c | c | c | c |}
\hline
 & Start & PLC 1 & PLC 2 & PLC 3 & PLC 4 & Finish \\
\hline
Start & 0 & 0.9 & 0.1 & 0 & 0 & 0 \\
\hline
PLC 1 & 0 & 0 & 0.9 & 0.1 & 0 & 0 \\
\hline
PLC 2 & 0 & 0 & 0 & 0.6 & 0.4 & 0 \\
\hline
PLC 3 & 0 & 0 & 0 & 0 & 0.9 & 0.1 \\
\hline
PLC 4 & 0 & 0 & 0 & 0 & 0 & 1 \\
\hline
Finish & 0 & 0 & 0 & 0 & 0 & 0 \\
\hline
\end{tabular}
\end{table}

\begin{figure}[htbp]
\centerline{\includegraphics[width=0.45\textwidth]{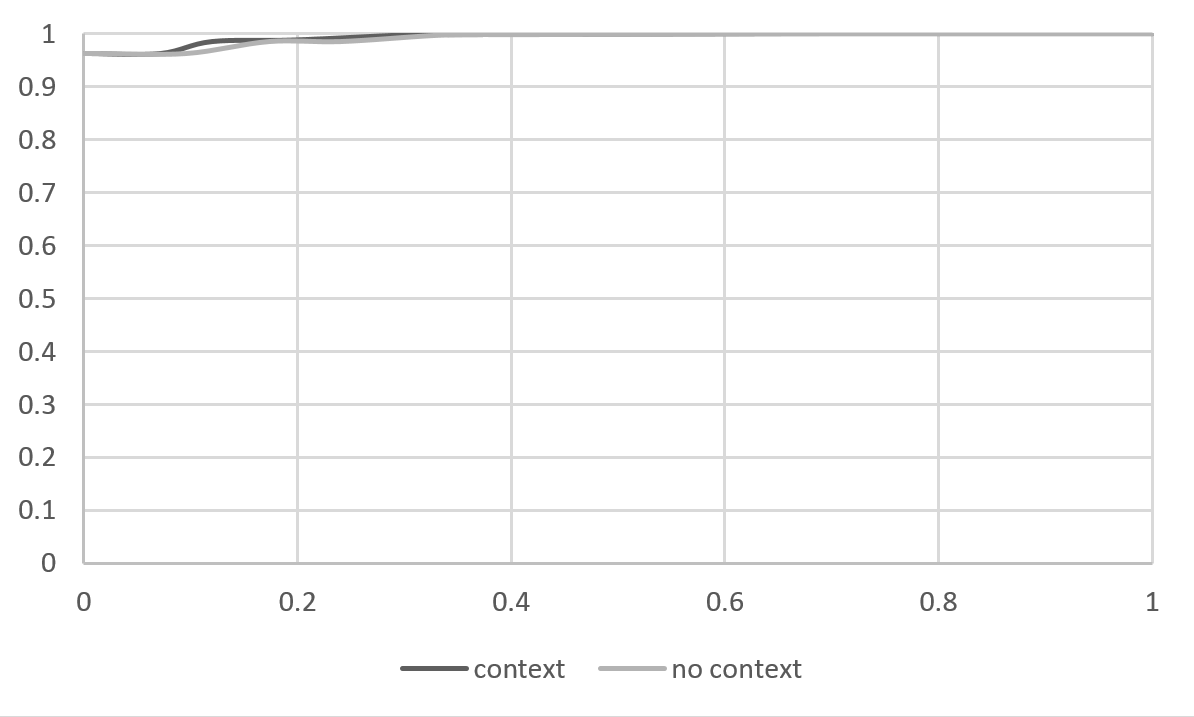}}
\caption{\acp{roc} for One Attack in Setup 1}
\label{fig:roc_setup1_static}
\end{figure}

The \acp{roc} for one malicious action in each sequence can be found in figure~\ref{fig:roc_setup1_static}.
It can be seen that they are 1 almost instantly.
This is due to the fact that the valid behaviour is relatively uniform,
making it easy to detect malicious activities.
For two and four malicious actions in each sequence,
the \ac{roc} starts even higher.

\begin{figure}[htbp]
\centerline{\includegraphics[width=0.45\textwidth]{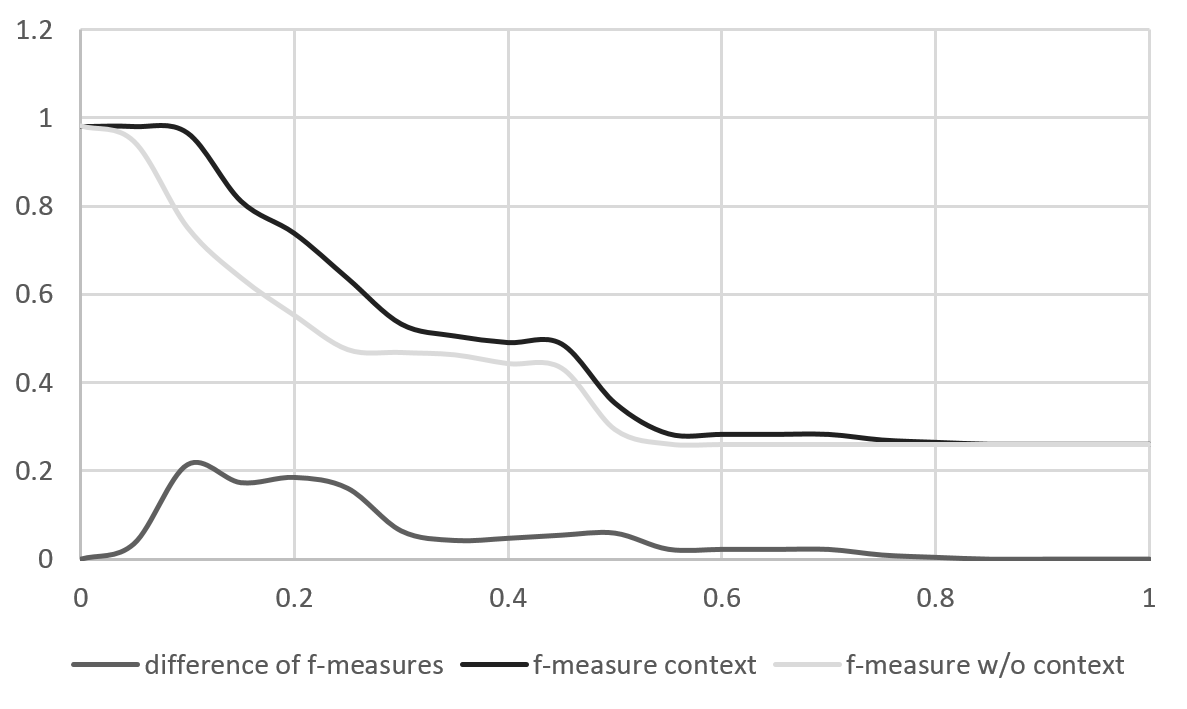}}
\caption{Difference of f-measures for One Attack in Setup 1}
\label{fig:f_setup1_static}
\end{figure}

In addition to the \ac{roc},
the f-measure is shown in figure~\ref{fig:f_setup1_static}.
The difference of f-measure for context-less and context-based approach is depicted as well.

Because of the higher number of false positives for the context-less approach,
the f-measure of the context-based approach is better in for certain thresholds.
With increasing threshold,
however,
the difference decreases as an increasing threshold leads to less transitions that fit.
The difference of f-measures looks similar for all numbers of malicious actions in this setup.

\paragraph{Setup 2}

\begin{figure}[htbp]
\centerline{\includegraphics[width=0.45\textwidth]{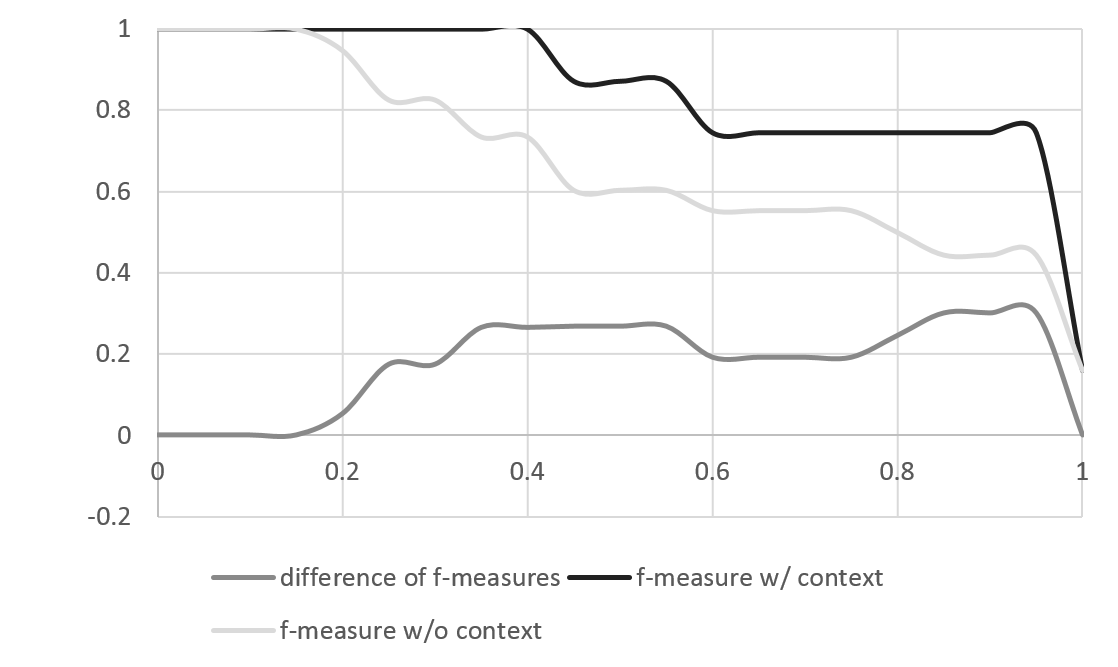}}
\caption{Difference of f-measures for One Attack in Setup 2}
\label{fig:f_setup2_static}
\end{figure}

In this setup,
the same \acp{ipc} as in the previous one are reconfigured when switching products.
They are summarised in table~\ref{tab:trans_probs}.
They,
however,
reprogram their corresponding \acp{plc} always in strict order.
The \ac{roc} immediately reaches close to one,
the diagram is therefore omitted.
The f-measures are shown in figure~\ref{fig:f_setup2_static}.
In the pictured scenario,
four attacks were introduced into every sequence.
The other scenarios look very much alike.

Sometimes,
the f-measure of the approach without context seems to be greater than one.
This is a result of plotting inaccuracy,
as the f-measure always results to a value between zero and one.
\\ \par
The strict order of valid actions makes it very easy to distinguish malicious from valid actions,
leading to such high true positive values of both approaches.
Furthermore,
the increase of the difference towards high thresholds results in the faster increase of false positive detections for the context-less approach.
While both methods perform very good in terms of true positives,
the context-based method is better with respect to choosing less false positives.

\paragraph{Setup 3}
In this setup,
each change of products led to reconfiguration of all \acp{ipc} and thus reprogramming of all \acp{plc}.
The \acp{plc} in turn are reprogrammed in strict numeric order.
However,
the order of reconfiguration varied,
depending on the product.
It can be found in table~\ref{tab:order_all_ipcs}.

\begin{table}
\centering
\caption{Order of Reconfiguring \acp{ipc} per Product}
\label{tab:order_all_ipcs}
\begin{tabular}{| l | c | c | c | c | c |}
\hline
 & first & second & third & fourth & fifth \\
 \hline
 Product 1 & IPC 1 & IPC 2 & IPC 3 & IPC 4 & IPC 5 \\
\hline
Product 2 & IPC 5 & IPC 4 & IPC 3 & IPC 1 & IPC 1 \\
\hline
Product 3 & IPC 1 & IPC 5 & IPC 2 & IPC 4 & IPC 3 \\
\hline
\end{tabular}
\end{table}

\begin{figure}[htbp]
\centerline{\includegraphics[width=0.45\textwidth]{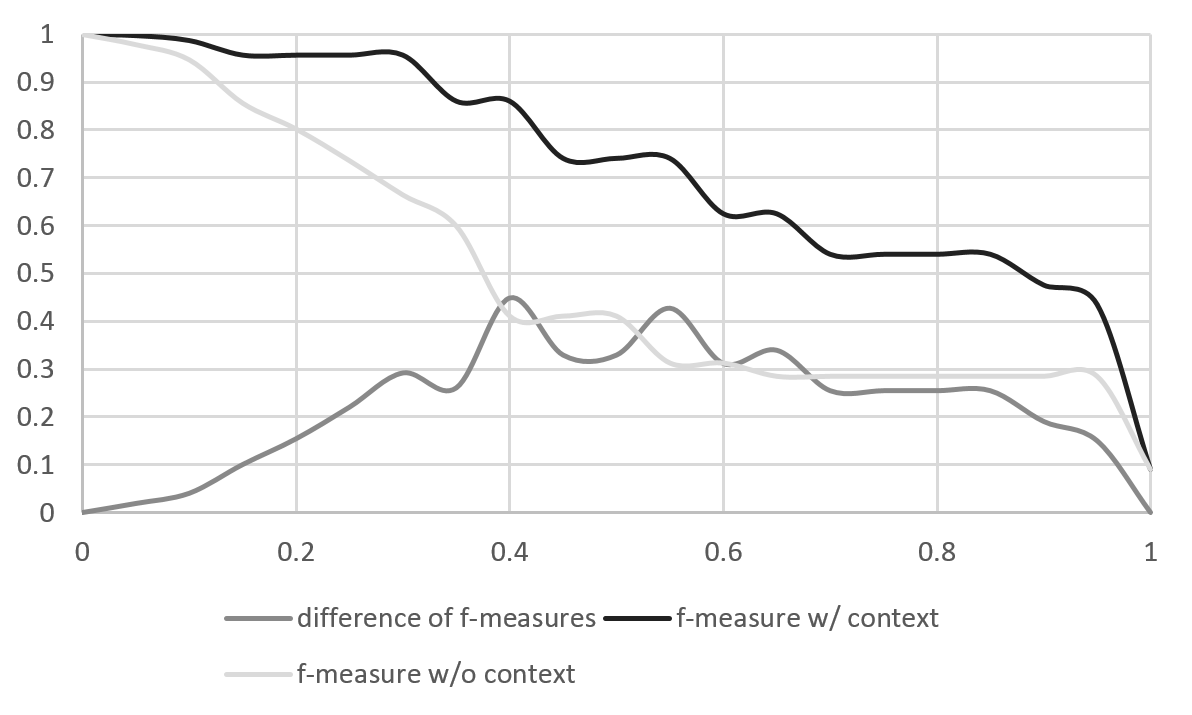}}
\caption{Difference of f-measures for One Attack in Setup 3}
\label{fig:f_setup3_static}
\end{figure}

Again,
the \ac{roc} starts with one due to the high true positive rate.
There is,
however,
a significant deviation in the f-measure,
as exemplarily shown in figure~\ref{fig:f_setup3_static} for four attacks per sequence.

The difference in the f-measures indicates an outperformance with respect to false positives by the context-based approach.

\begin{figure}[htbp]
\centerline{\includegraphics[width=0.45\textwidth]{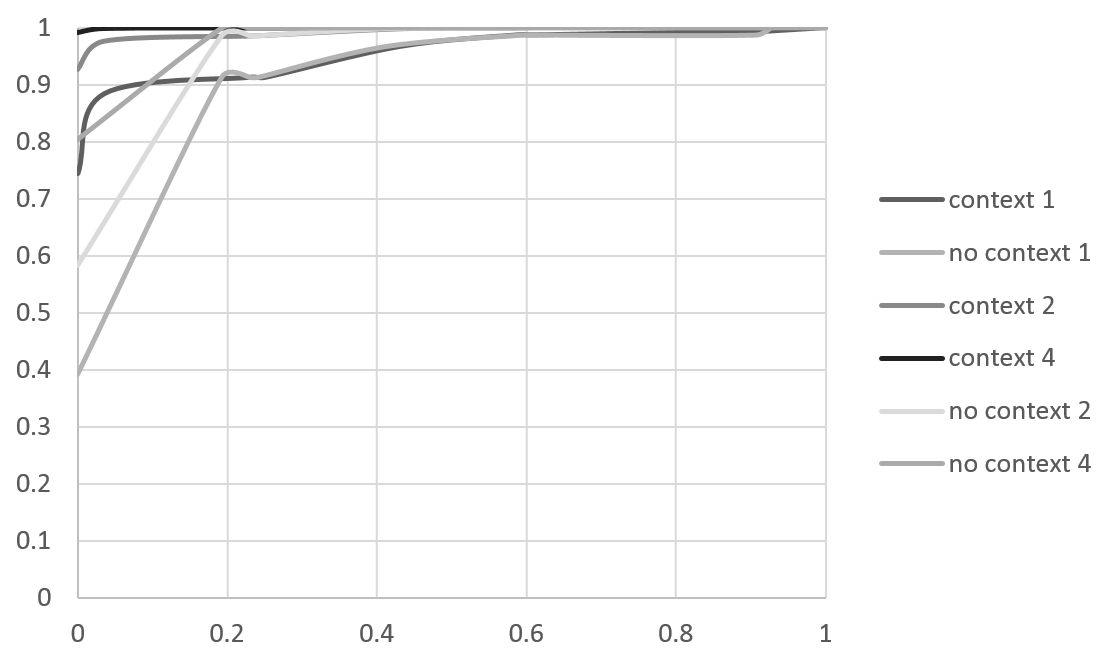}}
\caption{All \acp{roc} in Setup 4}
\label{fig:roc_setup4_static}
\end{figure}

\paragraph{Setup 4}
In this setup,
upon change of product,
each \ac{ipc} was reconfigured, 
as described in table~\ref{tab:order_all_ipcs}.
The order of reprogramming the \acp{plc} was completely random.
In figure~\ref{fig:roc_setup4_static},
the \acp{roc} for all numbers of malicious actions,
with and without the notion of context, 
respectively,
are drawn.
It can be seen that there is a break even point,
where the curves align.
Before that,
detection with the notion of context outperforms detection without consideration of context.

\begin{figure}[htbp]
\centerline{\includegraphics[width=0.45\textwidth]{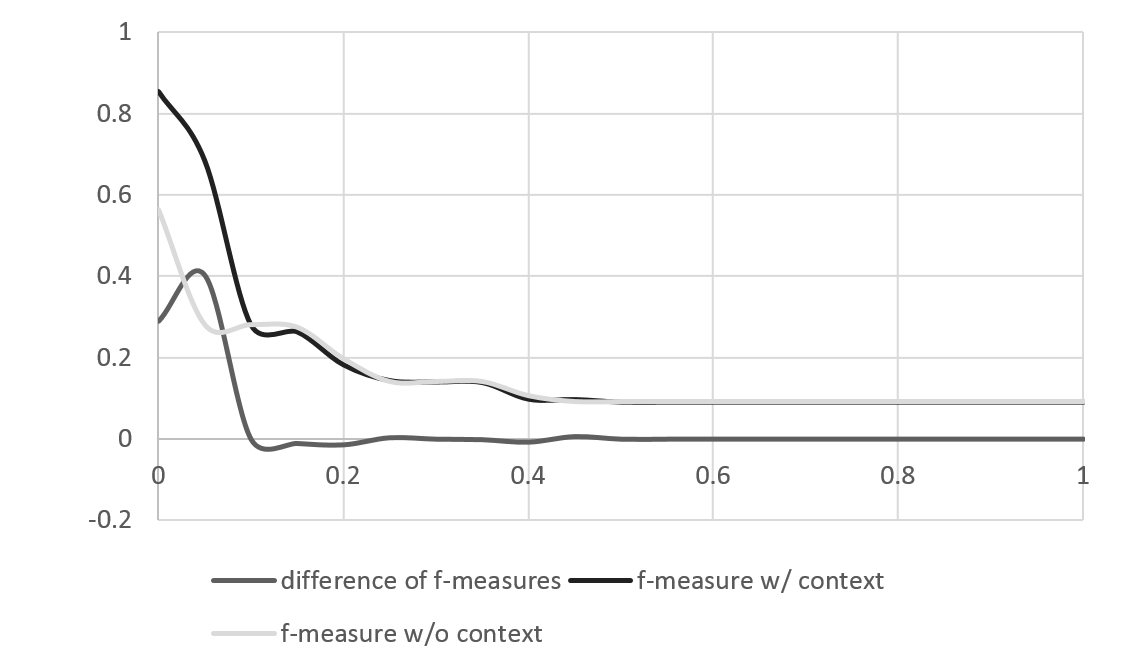}}
\caption{Difference of f-measures for One Attack in Setup 4}
\label{fig:f_setup4_static}
\end{figure}

The f-measures,
depicted in figure~\ref{fig:f_setup4_static} for one malicious action per valid sequence,
indicate a lower performance than in the previous setups. 
The difference is located in the low thresholds,
due to the better performance with respect to false positives of the context-based approach.
The comparatively worse performance is a result of the randomness of valid actions.

\subsection{Attacks in Random Order}
\label{ssec:airo}

In these setups,
the attacks always were in random order.
In turn,
one,
two and four attacks were introduced into four different kinds of valid actions.

\paragraph{Setup 1}

\begin{figure}[htbp]
\centerline{\includegraphics[width=0.45\textwidth]{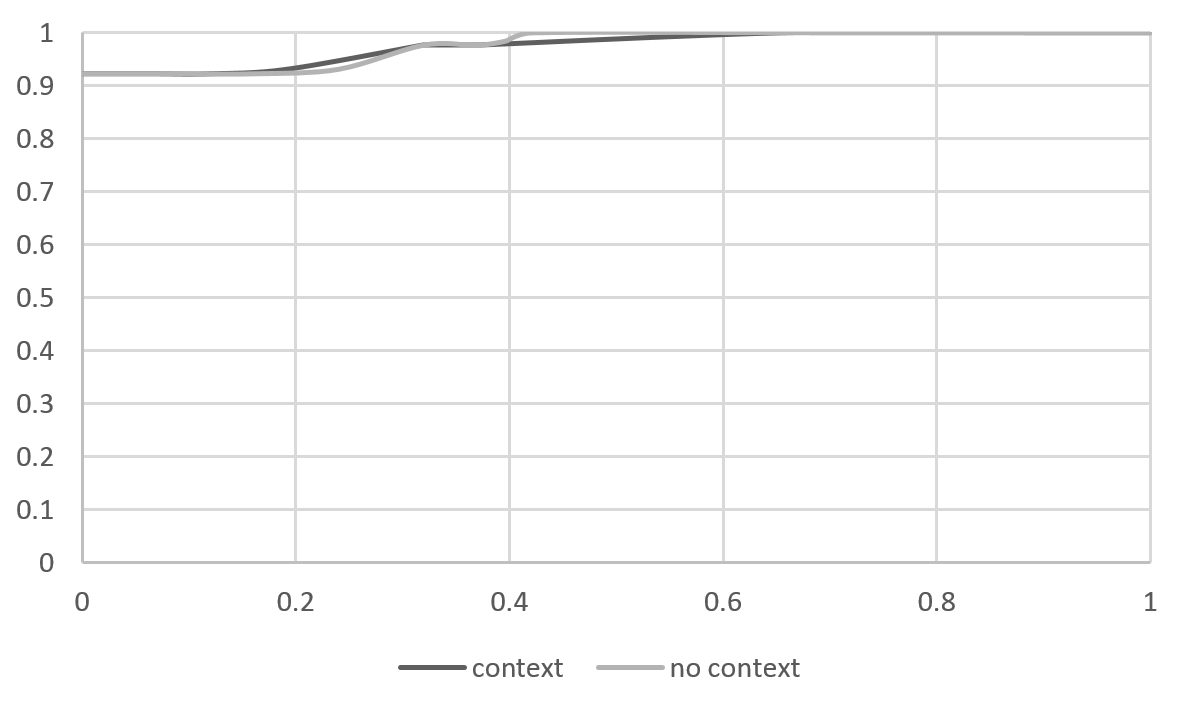}}
\caption{\acp{roc} for One Attack in Setup 1}
\label{fig:roc_setup1_rand}
\end{figure}

The reconfiguration pattern of \acp{ipc} is similar to the one described in table~\ref{tab:ipc_prods} of Setup 1 in subsection~\ref{ssec:aiso},
with product transition probabilities as described in table~\ref{tab:trans_probs}.
The reprogramming pattern of the \acp{plc} is shown in figure~\ref{fig:plc_trans}.
the according transition probabilities can be found in table~\ref{tab:plc_probs}.
The results are similar as well,
the \acp{roc} saturate almost from the beginning,
as can be seen in figure~\ref{fig:roc_setup1_rand}.

\begin{figure}[htbp]
\centerline{\includegraphics[width=0.45\textwidth]{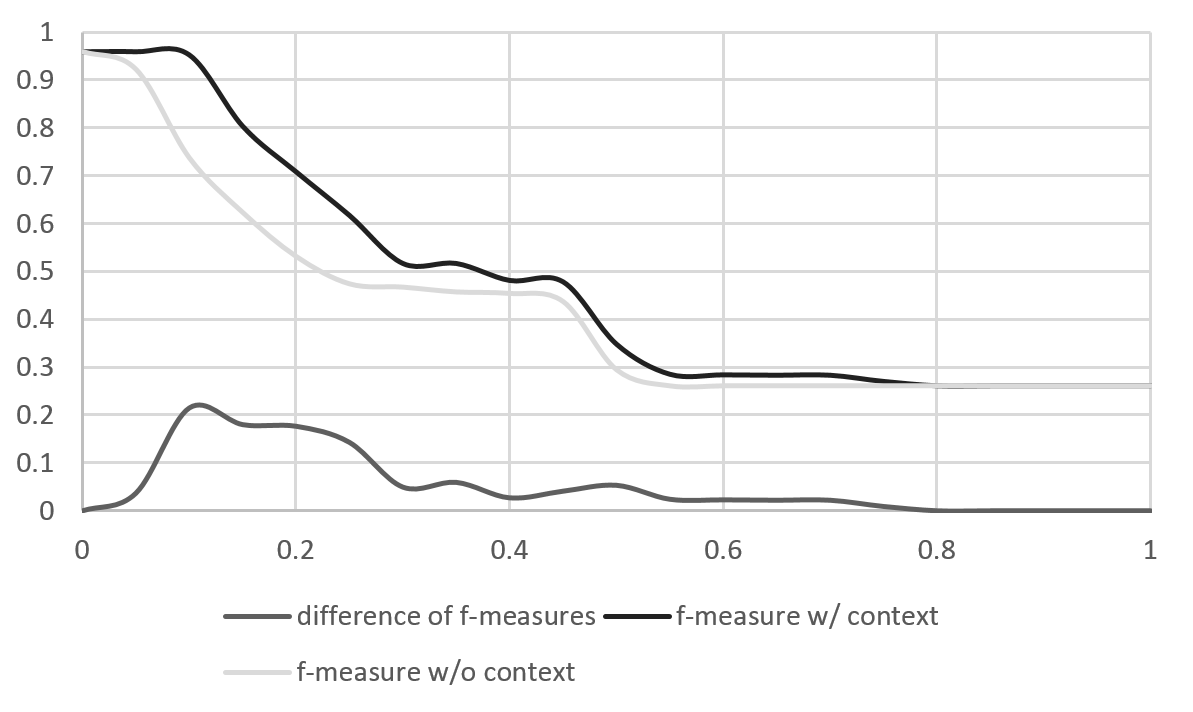}}
\caption{Difference of f-measures for One Attack in Setup 1}
\label{fig:f_setup1_rand}
\end{figure}

There is a small difference,
however,
in the f-measure of detection with and without the notion of context.
As shown in figure~\ref{fig:f_setup1_rand},
the detection without context slightly outperforms the detection without context in threshold ranges between $0.05$ and $0.3$.

\paragraph{Setup 2}

\begin{figure}[htbp]
\centerline{\includegraphics[width=0.45\textwidth]{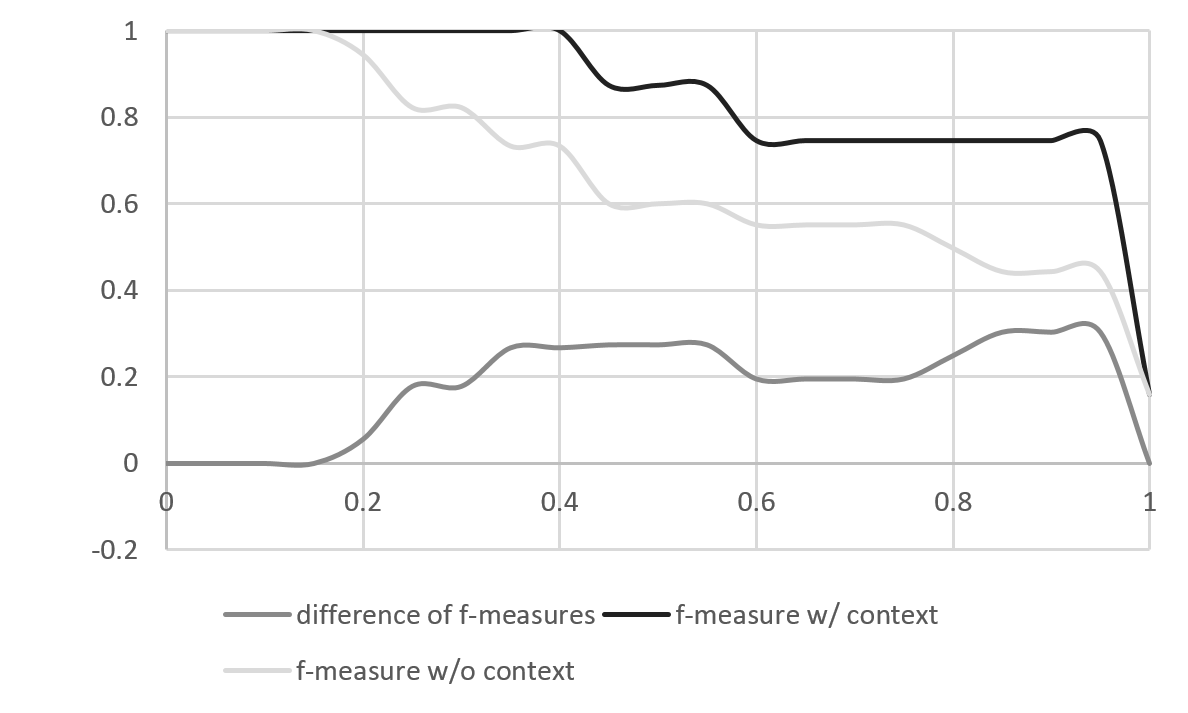}}
\caption{Difference of f-measures for Four Attacks in Setup 2}
\label{fig:f_setup2_rand}
\end{figure}

As in Setup 2 of subsection~\ref{ssec:aiso},
upon reconfiguration of an \ac{ipc},
according to tables~\ref{tab:ipc_prods} and~\ref{tab:trans_probs},
all its \acp{plc} are reprogrammed in strict order.
The \acp{roc} of one,
two and four attacks per sequence start with one,
depicting them is therefore of little interest.
Due to the false positives,
however,
the f-measure values differ,
especially with threshold greater than $0.2$.
This behaviour is shown in figure~\ref{fig:f_setup2_rand}.

\paragraph{Setup 3}

\begin{figure}[htbp]
\centerline{\includegraphics[width=0.45\textwidth]{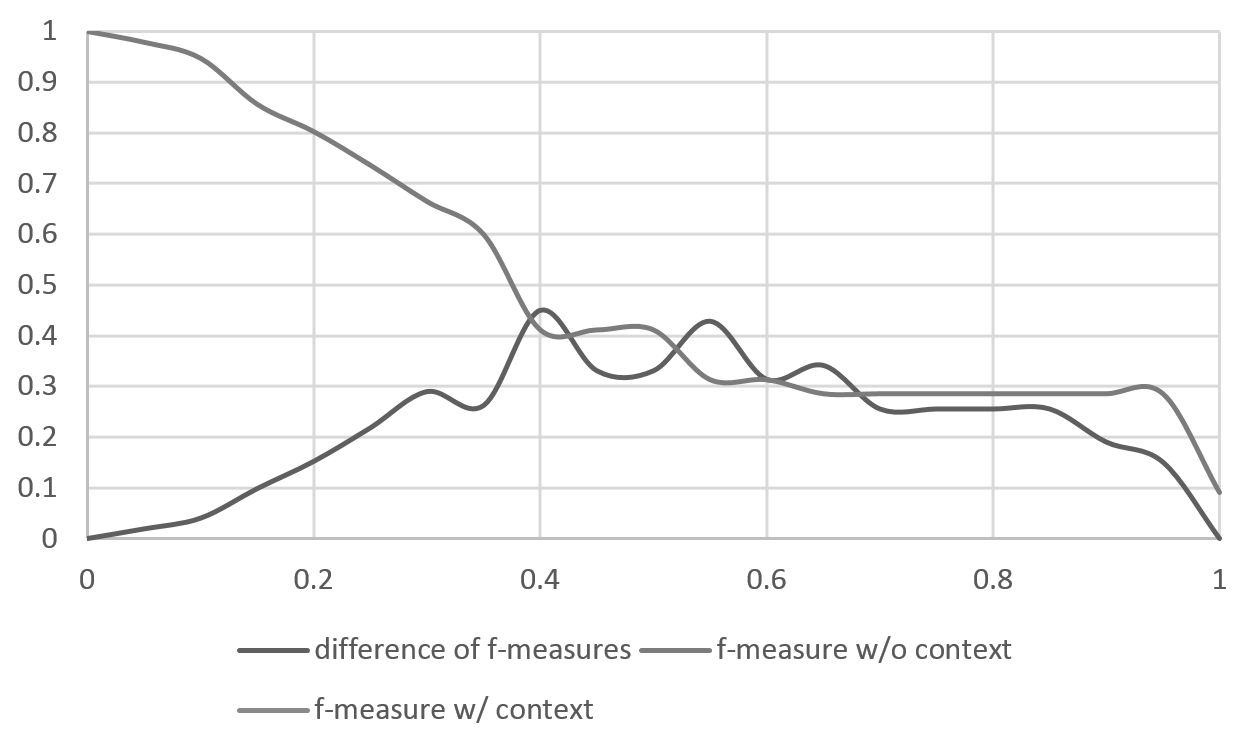}}
\caption{Difference of f-measures for One Attack in Setup 3}
\label{fig:f_setup3_rand}
\end{figure}

As with Setup 3 in subsection~\ref{ssec:aiso},
in this setup, 
all \acp{ipc} are reconfigured when the product is switched.
The order of reconfiguration for each product can be found in table~\ref{tab:order_all_ipcs}.
The corresponding \acp{plc} are reprogrammed in numeric order, 
respectively.
The \acp{roc} converge immediately.
There can,
however,
be found significant differences in the f-measures,
as shown in figure~\ref{fig:f_setup3_rand}.
Especially the area between $0.4$ and $0.6$ shows a difference due to false negatives,
% ST: especially the are??
% SDA: ... one letter, whole new meaning
where the detection with notion of context outperforms the detection without notion of context.

\paragraph{Setup 4}

\begin{figure}[htbp]
\centerline{\includegraphics[width=0.45\textwidth]{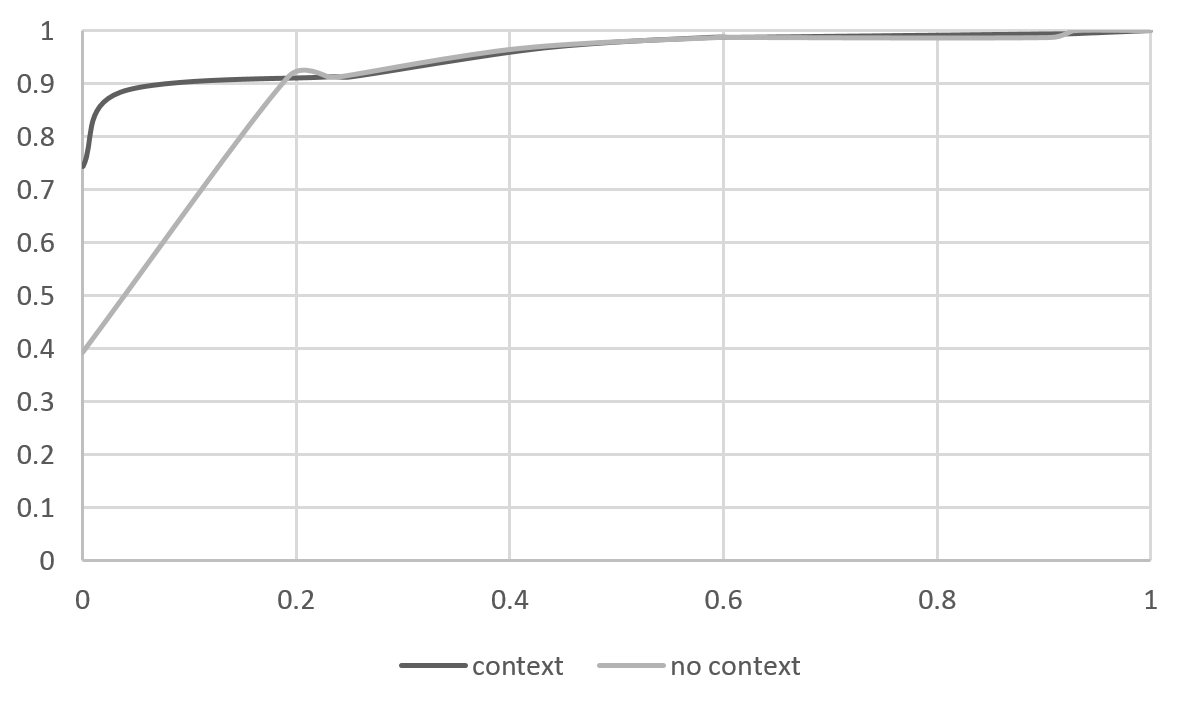}}
\caption{The \acp{roc} for One Attack per Sequence in Setup 4}
\label{fig:roc_setup4_rand}
\end{figure}

In this Setup,
the order of reconfiguration of \acp{ipc} again depends on the product as shown in table~\ref{tab:order_all_ipcs}.
The order of reprogramming the \acp{plc},
however,
is random for each \ac{ipc} respectively.
The \acp{roc} for one attack per sequence can be seen in figure~\ref{fig:roc_setup4_rand}.

\begin{figure}[htbp]
\centerline{\includegraphics[width=0.45\textwidth]{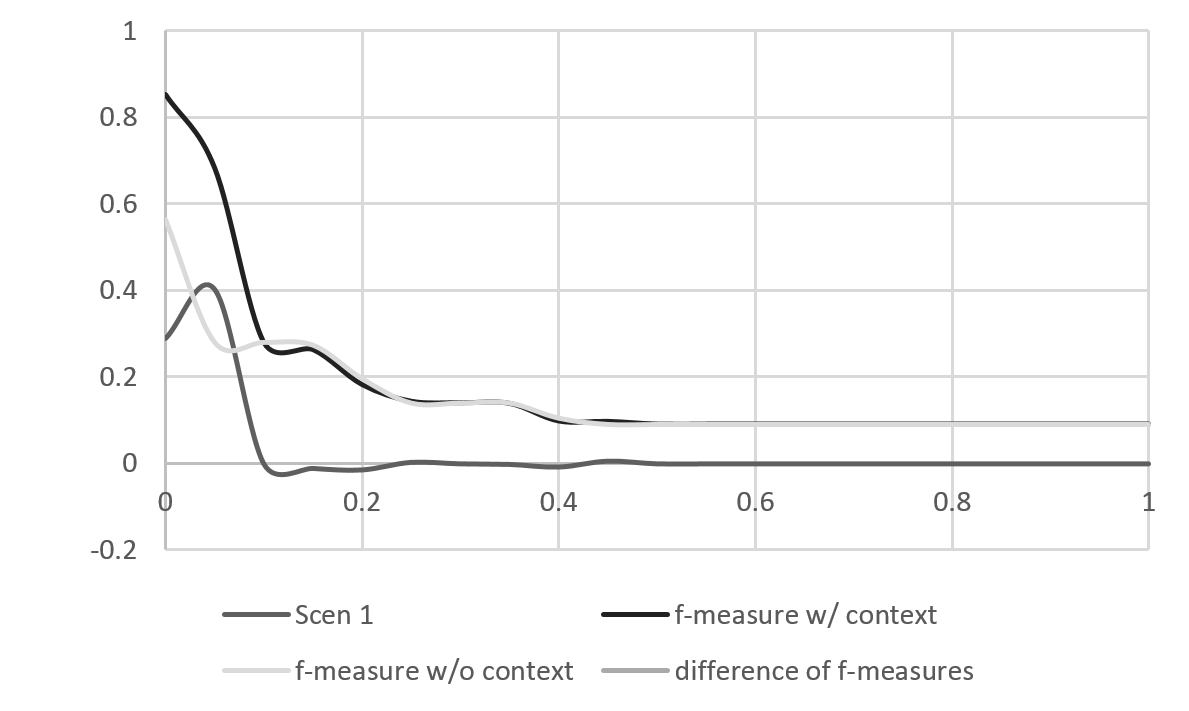}}
\caption{Difference of f-measures for One Attack in Setup 4}
\label{fig:f_setup4_rand}
\end{figure}

Up to a threshold value of $0.2$,
the detection with consideration of context performs significantly better.
% ST: considerating? oder consideration OF
This can also be seen in the f-measures, 
as depicted in figure~\ref{fig:f_setup4_rand}.

The \ac{roc} for detection without context seems to decrease in figure~\ref{fig:roc_setup4_rand}.
This is mathematically impossible and is due to an inaccuracy in plotting the graph.

\subsection{Results}
\label{ssec:results}
Throughout the experiments,
detection with consideration of context always performs better than detection without consideration of context.
This outperformance can in some cases be derived from the faster increasing \ac{roc}.
In any case,
the f-measure of context-based detection is higher and decreasing slower than the f-measure of non-context-based detection.
Therefore,
the false negative rate of context-based detection is lower than the false negative rate of context-less detection.
\\ \par
Furthermore,
it shows that the quality of detection,
in terms of \acp{roc},
depends on the randomness of the valid actions.
If only strict patterns occur,
detection is easy.
However,
if there is a level of uncertainty,
detection becomes more difficult,
as the transition matrices become more ambiguous.
\\ \par
Another interesting result lies in the discovery that the order of attacks has very low influence on the detection rate.
This is probably because of the fact that the transition from valid action to malicious action,
and vice versa,
serves as a distinguishing event.
Transitions within the malicious actions are not necessary to detect them.

\section{Conclusion}
\label{sec:conclusion}
Industrial systems and networks are typically more uniform and deterministic than classic internet networks.
This fact can be used to aid in intrusion detection.
If context information is considered in conjunction with network information,
intelligence about the presence or absence of malicious activities can be gained more easily.
Even though the proposed experiment was a simplification of a real industrial network,
the effect is noteworthy.
Especially false positives are significantly lower with than without notion of context.
In real-world applications,
this is a highly important factor,
as each false positive will lead to investigation and consume time and effort,
resulting in high costs.
A low false positive rate is therefore critical for acceptance of an detection algorithm.
\\ \par
For our future work,
we plan on creating a more sophisticated simulator on a packet base,
so that frequency,
content,
length and other properties of network traffic can be observed.
In addition to that,
we plan on integrating host-based information,
such as log-entries or settings of production tools,
into the detection solution.
This work suggests that this will lead to an increase in detection with a decrease in false positives.

\section*{Acknowledgment}
This work has been supported by the Federal Ministry of Education and Research of the Federal Republic of Germany (Foerderkennzeichen KIS4ITS0001, 16KIS0311, IUNO).
The authors alone are responsible for the content of the paper.

\bibliography{bib}{}

\begin{thebibliography}{10}

\bibitem{Igure.2006}
V.~M. Igure, S.~A. Laughter, and R.~D. Williams, ``Security issues in scada
  networks,'' {\em Computers {\&} Security}, vol.~25, pp.~498--506, 2006.

\bibitem{Zhou.2011}
S.~Zhou, J.~Han, and H.~Tang, ``Fractal traffic analysis and applications in
  industrial control ethernet network,'' in {\em Emerging Research in
  Artificial Intelligence and Computational Intelligence. AICI 2011.
  Communications in Computer and Information Science} (H.~Deng, D.~Miao,
  F.~Wang, and J.~Lei, eds.), vol.~237, Springer, Berlin, Heidelberg, 2011.

\bibitem{Zhu.2011}
B.~Zhu, A.~Joseph, and S.~Sastry, ``A taxonomy of cyber attacks on scada
  systems,'' {\em 2011 International Conference on Internet of Things and 4th
  International Conference on Cyber, Physical and Social Computing},
  pp.~380--388, 2011.

\bibitem{Porros.2010}
A.~Porros, ``Nuking and defending scada networks,'' 2010.

\bibitem{Luh.2017}
R.~Luh, S.~Marschalek, M.~Kaiser, H.~Janicke, and S.~Schrittwieser,
  ``Semantics-aware detection of targeted attacks: A survey,'' {\em Journal of
  Computer Virology and Hacking Techniques}, vol.~13, no.~1, pp.~47--85, 2017.

\bibitem{Roesch.1999}
M.~Roesch, ``Snort - lightweight intrusion detection for networks,'' in {\em
  Proceedings of the 13th USENIX Conference on System Administration}, LISA
  '99, (Berkeley, CA, USA), pp.~229--238, {USENIX Association}, 1999.

\bibitem{Paxson.1999}
V.~Paxson, ``Bro: A system for detecting network intruders in real-time,'' {\em
  Computer Networks}, vol.~31, no.~23-24, pp.~2435--2463, 1999.

\bibitem{JoshuaS.White.2013}
T.~F. N.~M. {Joshua S. White}, ``Quantitative analysis of intrusion detection
  systems: Snort and suricata,'' {\em Proc.SPIE}, vol.~8757, 2013.

\bibitem{Hindarto.2010}
D.~A. Hindarto, ``Wireless attacks from an intrusion detection perspective,''
  2010.

\bibitem{Jyothsna.2011}
V.~Jyothsna, V.~R. Prasad, and K.~M. Prasad, ``A review of anomaly based
  intrusion detection systems,'' {\em International Journal of Computer
  Applications}, vol.~28, no.~7, pp.~26--35, 2011.

\bibitem{Thames.2017}
L.~Thames and D.~Schaefer, eds., {\em Cybersecurity for Industry 4.0: Analysis
  for Design and Manufacturing}.
\newblock Springer Series in Advanced Manufacturing, Cham and s.l.: {Springer
  International Publishing}, 2017.

\bibitem{Ye.2000}
N.~Ye, ``A markov chain model of temporal behavior for anomaly detection,''
  2000.

\bibitem{Du.2004}
Y.~Du, H.~Wang, and Y.~Pang, ``Hmms for anomaly intrusion detection,'' in {\em
  Computational and Information Science} (D.~Hutchison, T.~Kanade, J.~Kittler,
  J.~M. Kleinberg, F.~Mattern, J.~C. Mitchell, M.~Naor, O.~Nierstrasz,
  C.~{Pandu Rangan}, B.~Steffen, M.~Sudan, D.~Terzopoulos, D.~Tygar, M.~Y.
  Vardi, G.~Weikum, J.~Zhang, J.-H. He, and Y.~Fu, eds.), vol.~3314 of {\em
  Lecture Notes in Computer Science}, pp.~692--697, Berlin, Heidelberg:
  {Springer Berlin Heidelberg}, 2004.

\bibitem{Hu.2009}
J.~Hu, X.~Yu, D.~Qiu, and H.-H. Chen, ``A simple and efficient hidden markov
  model scheme for host-based anomaly intrusion detection,'' {\em IEEE
  Network}, vol.~23, no.~1, pp.~42--47, 2009.

\bibitem{Yolacan.2016}
E.~N. Yola{\c{c}}an and D.~R. Kaeli, ``A framework for studying new approaches
  to anomaly detection,'' {\em International Journal of Information Security
  Science}, vol.~5, no.~2, pp.~39--50, 2016.

\bibitem{Chen.2016}
C.-M. Chen, D.-J. Guan, Y.-Z. Huang, and Y.-H. Ou, ``Anomaly network intrusion
  detection using hidden markov model,'' {\em International Journal of
  Innovative Computing, Information and Control,}, vol.~12, no.~2,
  pp.~569--580, 2016.

\bibitem{Zohrevand.2016}
Z.~Zohrevand, U.~Glasser, H.~Y. Shahir, M.~A. Tayebi, and R.~Costanzo, ``Hidden
  markov based anomaly detection for water supply systems,'' in {\em Big Data
  (Big Data), 2016 IEEE International Conference on}, pp.~1551--1560, IEEE,
  2016.

\bibitem{Fraunholz.2017}
D.~Fraunholz, S.~Duque~Ant\'{o}n, and H.~D. Schotten, ``Introducing gamfis: A
  generic attacker model for information security,'' in {\em Proceedings of the
  25th International Conference on Software, Telecommunications and Computer
  Networks. International Conference on Software, Telecommunications and
  Computer Networks (SoftCom-17), 25th, September 21-23, Split, Croatia}, IEEE,
  9 2017.

\bibitem{Northcutt.2005}
S.~Northcutt, L.~Zeltser, S.~Winters, K.~Kent, and R.~W. Ritchey, {\em Inside
  Network Perimeter Security (2Nd Edition) (Inside)}.
\newblock Indianapolis, IN, USA: Sams, 2005.

\bibitem{Dolev.1983}
D.~Dolev and A.~C. Yao, ``On the security of public key protocols,'' in {\em
  22nd Annual IEEE Symposium on Foundations of Computer Science}, pp.~198--208,
  IEEE, 2017.

\bibitem{Baum.1966}
L.~E. Baum and T.~Petrie, ``Statistical inference for probabilistic functions
  of finite state markov chains,'' {\em The Annals of Mathematical Statistics},
  pp.~1554--1563, 1966.

\bibitem{Eberle.2015}
A.~Eberle, ``Markov processes,'' March 2015.

\bibitem{Duqueanton.2017}
S.~Duque~Ant\'{o}n, D.~Fraunholz, J.~Zemitis, F.~Pohl, and H.~D. Schotten,
  ``Highly scalable and flexible model for effective aggregation of
  context-based data in generic iiot scenarios,'' in {\em 9th Central European
  Workshop on Services and their Composition. Central European Workshop on
  Services and their Composition (ZEUS-2017), February 13-14, Lugano,
  Switzerland} (O.~Kopp, J.~Lenhard, and C.~Pautasso, eds.), pp.~51--58, CEUR
  Workshop Proceedings, 4 2017.

\bibitem{Rijsbergen.1979}
C.~J. van Rijsbergen, ``Information retrieval,'' 1979.

\end{thebibliography}
\bibliographystyle{ieeetr}

\end{document}